\documentclass[a4paper]{revtex4}
\usepackage[english]{babel}
\usepackage{array,booktabs}
\usepackage{array} 
\usepackage{lipsum}   
\usepackage{calc}
\usepackage{pdflscape}
\usepackage{color}
\usepackage{float}
\usepackage[font=small,labelfont=bf]{caption}
\usepackage{graphicx}
\usepackage{amsmath}
\usepackage[nodisplayskipstretch]{setspace}

\usepackage{setspace}
\usepackage{tabularx}

\usepackage{float}
\usepackage{color}
\usepackage{amsmath}
\usepackage{float}
\usepackage{calc}
\usepackage{pdflscape}
\usepackage{color}
\usepackage{float}
\usepackage{subfigure}
\usepackage[font=small,labelfont=bf]{caption}
\usepackage{graphicx}
\begin{document}{\setlength\abovedisplayskip{4pt}}

\title{Octant Degeneracy, Quadrant of Leptonic CPV phase at Long Baseline $\nu$ Experiments and Baryogenesis}
\author{Kalpana Bora}
\affiliation{Department Of Physics, Gauhati University, Guwahati-781014, India}
\author{Gayatri Ghosh}
\affiliation{Department Of Physics, Gauhati University, Guwahati-781014, India}
\author{Debajyoti Dutta}
\affiliation{Harish-Chandra Research Institute, Chhatnag Road, Jhunsi, Allahabad 211019, India}


\begin{abstract}
In a recent work by us, we have studied, how CP violation discovery potential can be improved at long baseline neutrino experiments (LBNE/DUNE), by combining with its ND (near detector) and reactor experiments. In this work, we discuss how this study can be further analysed to resolve entanglement of the quadrant of leptonic CPV phase and Octant of atmospheric mixing angle $ \theta_{23} $, at LBNEs. The study is done for both NH (Normal hierarchy) and IH (Inverted hierarchy), HO (Higher Octant) and LO (Lower Octant). We show how baryogenesis can enhance the effect of resolving this entanglement, and how possible values of the leptonic CP-violating phase $ \delta_{CP} $  can be predicted
in this context. With respect to the latest global fit data of neutrino mixing angles, we predict the values of $ \delta_{CP} $  for different cases. In this context we present favoured values of $ \delta_{CP} $ ($ \delta_{CP} $ range at $ \geq $ 2$ \sigma $ ) constrained by the latest updated BAU range  and also confront our predictions of $ \delta_{CP} $  with an  up-to-date global analysis of neutrino oscillation data. We find that some region of the favoured  $ \delta_{CP} $ parameter space lies within the best fit values  around  $ \delta_{CP} \simeq 1.3\pi-1.4 \pi $.  A detailed analytic and numerical study of baryogenesis through leptogenesis is performed in this framework in a model independent way. 
\end{abstract}

\maketitle
\section{Introduction}	
\setlength{\baselineskip}{13pt}
Today, physics is going through precision era-this is more so for Neutrino physics. With the measurement of reactor mixing angle $\theta_{13}$ \cite{Fgli,Frero,Garc} precisely by reactor experiments, the unknown quantities left to be measured in neutrino sector are $-$ leptonic CP violating phase \cite{DK,MG,PT,Kang,LHCb,Patrik}, octant of atmospheric angle $\theta_{23}$ \cite{KD,Gonzalez,Animesh,Choubey,Daljeet}, mass hierarchy, nature of neutrino etc. Long baseline neutrino experiments (LBNE \cite{LBNE,Akiri},  NO$ \nu $A \cite{Ayres} , T2K \cite{T2K}, MINOS \cite{Minos}, LBNO \cite{DA} etc) may be very promising, in measuring many of these sensitive parameters.
\par 
Measuring leptonic CP violation (CPV) is one of the most demanding tasks in future
neutrino experiments \cite{Branco}. The relatively large value of the reactor mixing angle $ \theta_{13} $ measured with a high precision in neutrino experiments \cite{Fgli} has opened up a wide range of possibilities to examine CP violation in the lepton sector. The leptonic CPV phase can be induced by the PMNS
neutrino mixing matrix \cite{Pcarvo} which holds, in addition to the three mixing angles, a Dirac type CP violating phase in general as it exists in the quark sector, and two extra phases if neutrinos are Majorana particles. Even if we do not yet have significant evidence for leptonic CPV, the current global fit to available neutrino data manifests nontrivial values of the Dirac-type CP phase \cite{Capp,Gon}. In this context, possible size of leptonic CP violation detectable through neutrino oscillations can be predicted. Recently, \cite{DK}, we have explored possibiities of improving CP violation discovery potential of newly planned Long-Baseline Neutrino Experiments (earlier LBNE, now called DUNE) in USA. In neutrino oscillation probability expression P($ \nu_{\mu}\rightarrow \nu_{e}$) relevant for LBNEs, the term due to significant matter effect, changes sign when oscillation is changed from neutrino to antineutrino mode, or vice-versa. Therefore in presence of matter effects, CPV effect is entangled and hence, one has two degenerate solutions - one due to CPV phase and another due to its entangled value. It has been suggested to resolve this issue by combining two experiments with different baselines \cite{Varger,Minakata}. But CPV phase measurements depends on value of reactor angle $\theta_{13}$, and hence precise measurement of $\theta_{13} $ plays crucial role in its CPV measurements. This fact was utilised recently by us \cite{DK}, where we have explored different possibilities of improving CPV sensitivity for LBNE, USA. We did so by considering LBNE with \\
1. Its ND (near detector).\\
2. And reactor experiments.
\par 
We considered both appearance ($ \nu_{\mu}\rightarrow \nu_{e}$) and disappearance ($ \nu_{\mu}\rightarrow \nu_{e}$) channels in both neutrino and antineutrino modes. Some of the observations made in \cite{DK} are\\
1. CPV discovery potential of LBNE increases significantly when combined with near detector and reactor experiments.\\ 
2. CPV violation sensitivity is more in LO (lower octant) of atmospheric angle $\theta_{23}$, for any assumed true hierarchy.\\
3. CPV sensitivity increases with mass of FD (far detector).\\
4. When NH is true hierarchy, adding data from reactors to LBNE improve its CPV sensitivity irrespective of octant.
\par 
Aim of this work is to critically analyse the results presented in \cite{DK}, in context of entanglement of quadrant of CPV phase and octant of $\theta_{23} $, and hence study the role of baryogenesis in resolving this enganglement. Though in \cite{DK}, we studied effect of both ND and reactor experiments on CPV sensitivity of the LBNEs, in this work we have considered only the effect
 of ND. But similar studies can also be done for the effect of Reactor experiments on LBNEs as well. The details of LBNE and ND are same
 as in \cite{DK}. Following the results of \cite{DK}, either of the two octants is favoured, and the enhancement of CPV sensitivity with
respect to its quadrant is utilized here to calculate the values of lepton-antilepton symmetry. This is done considering two cases of the
rotation matrix for the fermions - CKM only, and CKM+PMNS. Then, this is used to calculate the 
value of BAU. 
\par 
This is an era of precision measurements in neutrino physics. We therefore consider variation of $\Delta m^{2}_{31}$ range 
at its 3$\sigma$ C.L Vs $ \delta_{CP} $ range at $ \geq $ 2$ \sigma $  over the corresponding distribution of $ \chi^{2} $-minima from figure 2. We calculate baryon to photon ratio, and compare with its experimentally known best fit value. As, constrained by the latest updated BAU limits, $5.7\times 10^{-10} < BAU < 6.7 \times 10^{-10} $,  we  plot $\theta_{13}$ range at its 3 $ \sigma $ C.L \cite{Frero} from its central value  Vs $ \delta_{CP} $ range at $ \geq $ 2$ \sigma $  over $ \chi^{2} $-minima distribution and find that for IH, LO case the allowed $ \theta_{13} $ has a varied range altering within $ 8.7^{0} $ to $ 9.4^{0} $ in the upper quadrant ($ \pi $ to $2\pi$) and $ 8.7^{0} $ to $ 9.8^{0} $ in the lower quadrant (0 to $ \pi $). Similarly for IH, HO case the allowed $ \theta_{13} $ has a varied range differing within $ 8.7^{0} $ to $ 9.8^{0} $. As shown in our results in Section IV, in IH, LO case the spectrum of $ \delta_{CP} $ is mostly concentrated in the region $ 88^{0} $ for $ \theta_{13} $ around  $ 9.09^{0} $ to $ 9.2^{0} $, $ 9.25^{0} $, $ 9.35^{0} $ to $ 9.5^{0} $, $ 9.6^{0} $ to $ 9.65^{0} $. Also   $ \delta_{CP} = 276.5^{0} $ exists for $\theta_{13}$ around $ 9.06^{0} $ to $ 9.12^{0} $, $ 9.3^{0} $ to $ 9.45^{0} $, $ 9.517^{0} $ and $ \delta_{CP} =  290^{0} $ for $\theta_{13}$ around , $ 9^{0} $, $ 9.15^{0} $ to $ 9.3^{0} $, $ 9.35^{0} $ to $ 9.5^{0} $ in the higher quadrant ($ \pi $ to 2$ \pi $) for $\theta_{13}$ around $ 9.09^{0} $ to $ 9.5^{0} $. Similarly as shown in our results in Sec. 4, as allowed by the updated BAU limits in IH, HO case, the parameter space of $ \delta_{CP} $ in the lower quadrant (0 to $\pi$) demands $ \delta_{CP}$ to be around $ 69^{0} $,  for $ \theta_{13}$ $\sim$ $8.86^{0}-8.896^{0}$, $9.06^{0}-9.22^{0}$, $9.32^{0}$, $9.61^{0}$. There also exists $ \delta_{CP} = 95^{0}$ which constrains $ \theta_{13} $ to be around   $ 8.99^{0} $, $ 9.24^{0} $, $ 9.352^{0} $, $ 9.5-9.64^{0} $ and $ 9.79^{0} $. In the upper quadrant ($ \pi $ to 2$ \pi $) for the present updated BAU constraint, the allowed region of $ \delta_{CP} $ parameter space becomes constrained with $ \delta_{CP} = 257.5^{0} $, for $ \theta_{13} $ around  $ 8.96^{0} $,  $ 9.0974^{0} $ to  $ 9.22^{0} $,  $ 9.5^{0} $,  $ 9.6^{0} $ and  $ 9.74^{0} $. Also the BAU constraint requires $ \delta_{CP} $ to be equal to $ 288^{0} $, for $\theta_{13}$ around  $ 9^{0} $, $ 9.06^{0} $ to $ 9.35^{0} $, $ 9.5^{0} $,  $ 9.69^{0} $,  $ 9.8 $. Also $ \delta_{CP} = 295^{0} $ survives for $\theta_{13}$ around  $ 9.09^{0} $, $ 9.26^{0} $, $ 9.34^{0} $, $ 9.5^{0} $,  $ 9.6^{0} $,  $ 9.69 $. A part of the allowed $ \delta_{CP} $ parameter space is found to lie within the best fit values  of $ \delta_{CP} \simeq 1.3\pi-1.4 \pi$.  As constrained by the current BAU bounds we present the 3-D variation of the favoured range of parameters: $\Delta m^{2}_{31}$ range within its 3$\sigma$ C.L, $\theta_{13}$ range 
at its 3$\sigma$ C.L and $\delta_{CP}$ range at $ \geq $ 2$ \sigma $ varied within $ [0,2\pi] $ in Fig. 2.
\par 
As can be seen from the results presented in Sect. IV, we find that, BAU can be explained most favourably for the following possible cases: $\delta_{CP}= 1.536 \pi$, IH, LO of $ \theta_{23} $; $\delta_{CP}= 1.611 \pi$, IH, LO of $ \theta_{23} $; $\delta_{CP}= 0.488 \pi$, IH, LO of $ \theta_{23} $; $\delta_{CP}= 1.638 \pi$, IH, HO of $ \theta_{23} $; $\delta_{CP}= 1.6 \pi$, IH, HO of $ \theta_{23} $; $\delta_{CP}= 1.43 \pi$, IH, HO of $ \theta_{23} $;   $\delta_{CP}= .777 \pi$, IH, HO of $ \theta_{23} $; $\delta_{CP}= .5277 \pi$, IH, HO of $ \theta_{23} $; $\delta_{CP}= 1.436 \pi$, NH, HO of $ \theta_{23} $. It is worth mentioning that the value of $\delta_{CP} = 1.43 \pi$ and $\delta_{CP} = 1.436 \pi$  is close to the central value of $ \delta_{CP} $ from the recent global fit result \cite{Gon, kol}. It is fascinating to notice that a nearly maximal CP-violating phase $\delta_{CP} = \frac{3}{2}\pi$ has been reported by
the T2K \cite{Abe}, NO$\nu$A \cite{Bian} and Super-Kamiokande experiments \cite{S}, even if the statistical significance of all these experimental results is below 3$\sigma$ level. This accords with one of our calculated favoured solution $\delta_{CP} = 1.536 \pi  $ which exactly holds with the current BAU constraints. Moreover, such hints of a nonzero $\delta_{CP} $  were already present in global analyses of neutrino oscillation data, such as the one in Ref. \cite{Capp}. Our main aim in this work is to carry out a detailed analysis of the breaking of the entanglement of the quadrant of leptonic CPV phase and Octant of $ \theta_{23} $ by using the current data of $ \nu $ mixing parameters and identify the CPV phase and $ \theta_{13} $ spectrum required to get the breaking favourable with the current BAU constraint. These results could be important keeping in view that the quadrant of leptonic CPV phase, and octant of atmospheric mixing angle $\theta_{23}$ are yet not fixed. Also, they are significant in context of precision measurements on neutrino oscillation parameters. 

\par
The paper is organized as follows. In Section II, we discuss entanglement of quadrant of  CPV phase and octant of $ \theta_{23} $. In Section III, we present a review on leptogenesis and baryogenesis. In Sec. IV we show how the  
baryon asymmetry (BAU) within the SO(10) model, by using two distinct forms for the lepton CP asymmetry, can be used to break the entanglement. Finally in Sec. V, we present our conclusions.

\section{CPV Phase and Octant of $\theta_{23}$}

As discussed above, from Fig. 3 of \cite{DK}, we find that by combining with ND and reactor experiments, CPV sensitivity of LBNE 
improves more for LO (lower octant) than HO (higher octant), for any assumed true hierarchy. In Fig. 1 below we plot CP asymmetry,
\begin{equation}
A_{CP} = \frac{P(\nu_{\mu}\rightarrow \nu_{e})-P(\overline{\nu_{\mu}}\rightarrow \overline{\nu_{e}})}{P( \nu_{\mu}\rightarrow \nu_{e})+P(\overline{\nu_{\mu}}\rightarrow \overline{\nu_{e}})}
\label{diseqn}
\end{equation}
as a function of leptonic CPV phase $ \delta_{CP} $, for 0 $\leq \delta_{CP} \leq 2 \pi$.
 CP asymmetry also depends on the mass hierarchy.
For NH, CP asymmetry is more in LO than in HO. For IH, CP asymmetry is more in LO than in HO. In this work we have used above information to calculate dependance of leptogenesis on octant of $ \theta_{23}$ and quadrant of CPV phase.
\begin{figure}[b]
\centerline{\includegraphics[width=7.8cm]{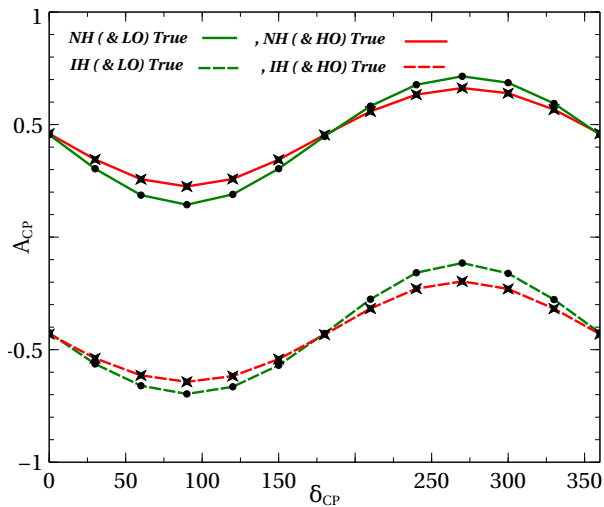}}
\caption{CP asymmetry vs $\delta_{CP}$ at DUNE/LBNE, for both the hierarchies. In Fig. 1 red and green solid (dotted) lines are for NH (IH) with types of curve to distinguish HO and LO as the true octant respectively.\label{pcv.pdf}}
\end{figure}
From Fig. 1 we see that
\par  
\begin{equation}
A_{CP}(LO) > A_{CP}(HO)
\label{diseqn}
\end{equation}
\par
For a given true hierarchy, there are eight degenerate solutions    
$$\delta_{CP}(\text{first quadrant}) - \theta_{23}(\text{lower octant})$$
$$\delta_{CP}(\text{second quadrant}) - \theta_{23}(\text{lower octant})$$
$$\delta_{CP}(\text{third quadrant}) - \theta_{23}(\text{lower octant})$$
$$\delta_{CP}(\text{fourth quadrant}) - \theta_{23}(\text{lower octant})$$
$$\delta_{CP}(\text{first quadrant}) - \theta_{23}(\text{higher octant})$$
$$\delta_{CP}(\text{second quadrant}) - \theta_{23}(\text{higher octant})$$
$$\delta_{CP}(\text{third quadrant}) - \theta_{23}(\text{higher octant})$$
\begin{equation}
\delta_{CP}(\text{fourth quadrant}) - \theta_{23}(\text{higher octant})
\label{diseqn}
\end{equation}
This eight-fold degeneracy can be viewed as 
\begin{equation}
\text{Quadrant of CPV phase} - \text{Octant of}\hspace{.1cm}  \theta_{23}
\label{diseqn}
\end{equation}
entanglement. Out of these eight degenerate solutions, only one should be true solution. To pinpoint one true solution, this entanglement has to be broken. We have shown \cite{DK} that sensitivity to discovery potential of CPV at LBNEs in LO is improved more, if data from near detector of LBNEs, or from Reactor experiments is added to data from FD of LBNEs as shown in Fig. 3 of \cite{DK}. Therefore 8-fold degeneracy of (3) gets reduced to 4-fold degeneracy, with our proposal \cite{DK}. Hence, following this 4-fold degeneracy still remains to be resolved.
 \begin{equation*}
\delta_{CP}(\text{first quadrant}) - \theta_{23}(\text{LO})
\end{equation*}
\begin{equation*}
\delta_{CP}(\text{second quadrant}) - \theta_{23}(\text{LO})
\end{equation*}
\begin{equation*}
\delta_{CP}(\text{third quadrant}) - \theta_{23}(\text{LO})
\end{equation*}
\begin{equation}
\delta_{CP}(\text{fourth quadrant}) - \theta_{23}(\text{LO})
\end{equation}

The possibility of $ \theta_{23} > 45^{0}$, ie HO of $ \theta_{23}$ is also considered in this work. In this context the degeneracy is
\begin{equation*}
\delta_{CP}(\text{first quadrant}) - \theta_{23}(\text{HO})
\end{equation*}
\begin{equation*}
\delta_{CP}(\text{second quadrant}) - \theta_{23}(\text{HO})
\end{equation*}
\begin{equation*}
\delta_{CP}(\text{third quadrant}) - \theta_{23}(\text{HO})
\end{equation*}
\begin{equation}
\delta_{CP}(\text{fourth quadrant}) - \theta_{23}(\text{HO})
\end{equation}
 
\begin{figure}[b]
\centerline{\begin{subfigure}[]{\includegraphics[width=6.8cm]{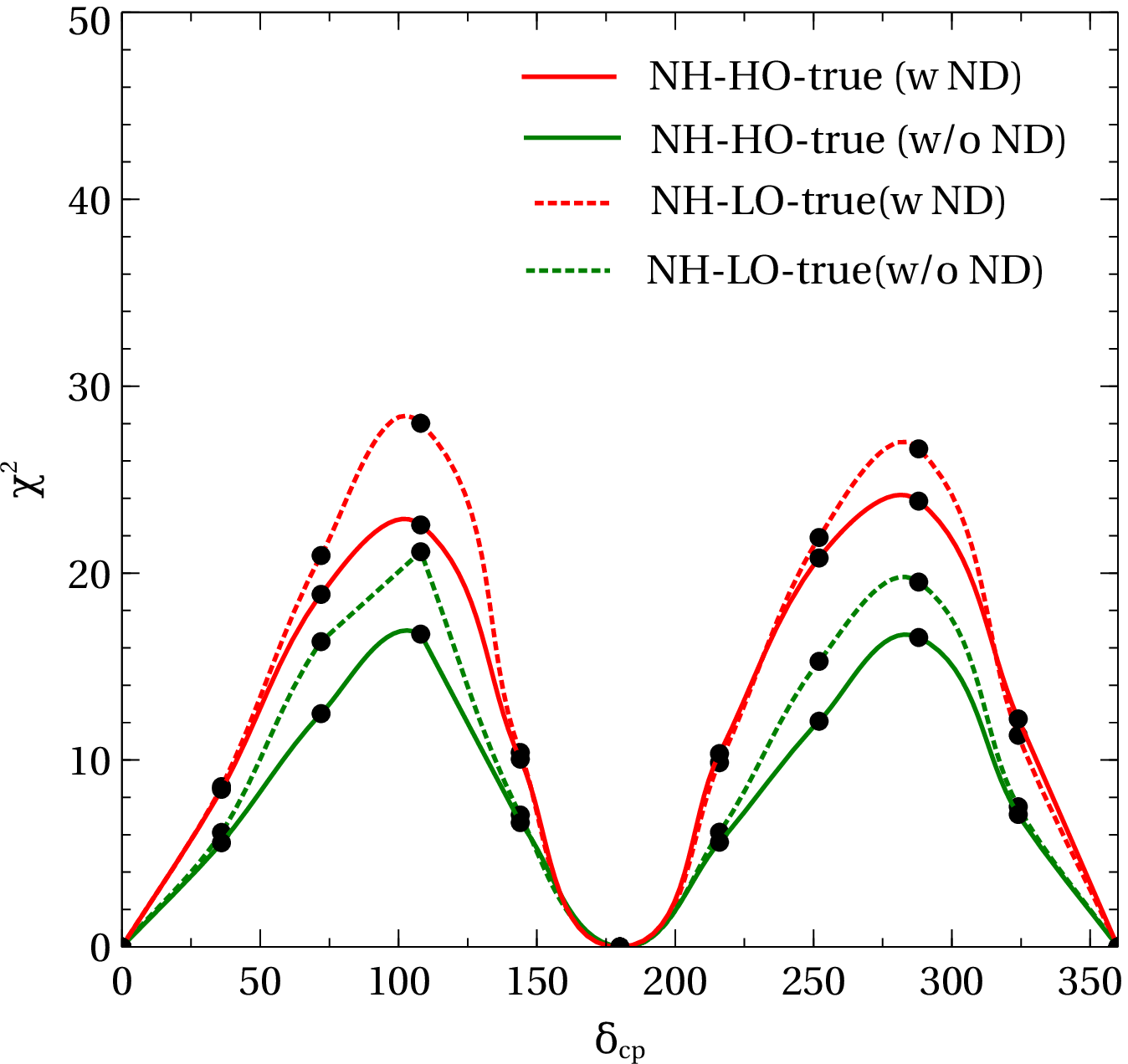}}\end{subfigure}
\begin{subfigure}[]{\includegraphics[width=6.8cm]{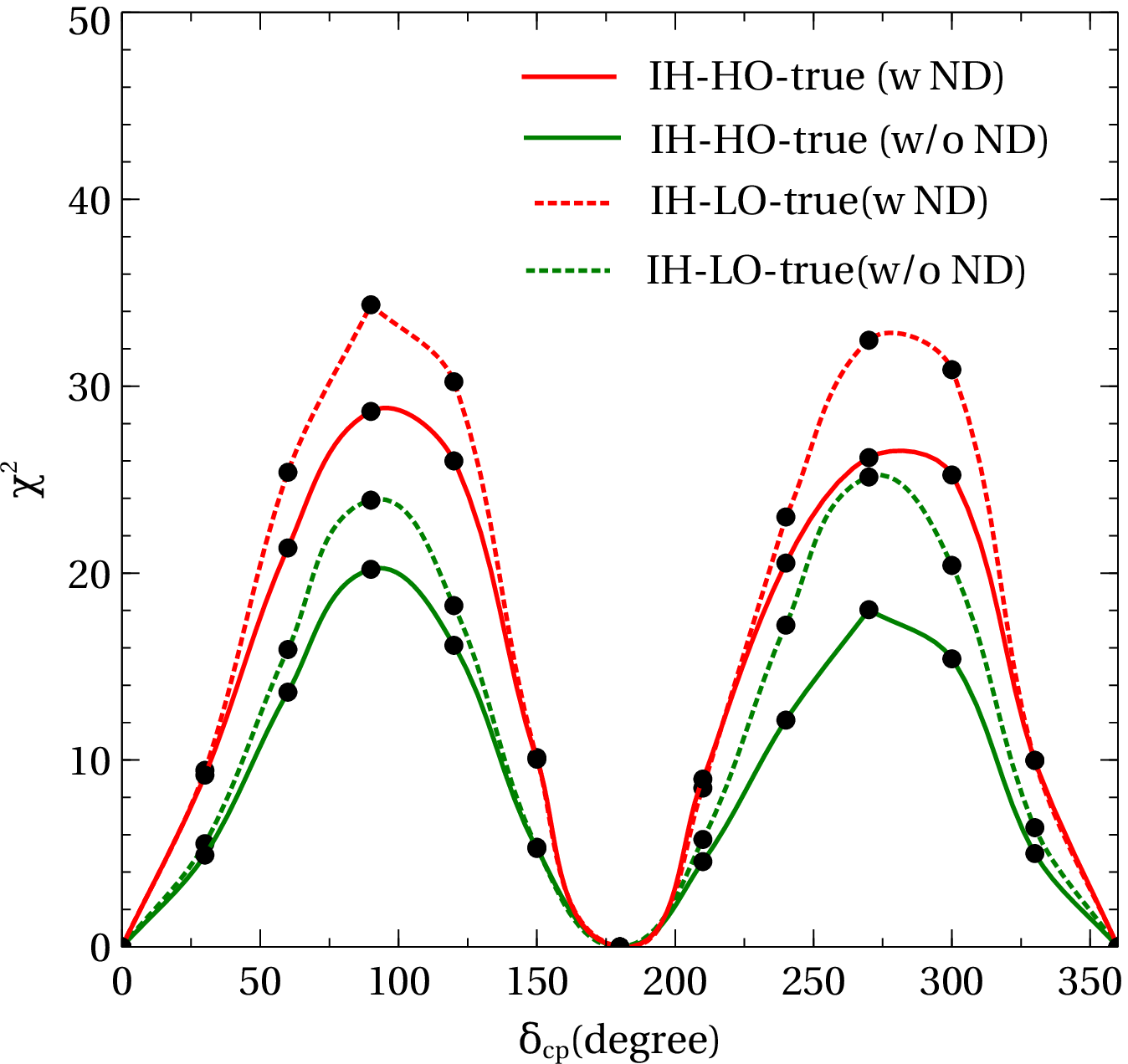}}\end{subfigure}}
\caption{In Fig. 2a and 2b $\delta_{CP}$ Vs $\chi^{2}$ sensitivity corresponding to CP discovery potential at LBNEs, for both the hierarchies and octant is shown.}
\end{figure}

In this work, we propose that leptogenesis can be used to break above mentioned 4-fold degeneracy of Eq. (5),(6). It is known that observed baryon asymmetry of the Universe (BAU) can be explained via leptogenesis \cite{Bari, RM, Bhupal, Pd, GC}. In leptogenesis, the lepton-antilepton asymmetry can be explained, if there are complex Yukawa couplings or complex fermion mass matrices. This in turn arises due to complex leptonic CPV phases, $\delta_{CP}$, in fermion mass matrices. If all other parameters except leptonic $\delta_{CP}$ phase in the formula for lepton - antilepton asymmetry are fixed, for example, then observed value of BAU from experimental observation can be used to constrain quadrant of $\delta_{CP}$, and hence 4-fold entanglement of (5),(6) can be broken. An experimental signature of CP violation associated to the Dirac phase $\delta_{CP}$, in PMNS matrix \cite{Mki}, can in principle be obtained, by searching for CP asymmetry in $ \nu $ flavor oscillation. To elucidate this proposal, we consider model independent scenario, in which BAU arises due to leptogenesis, and this lepton-antilepton asymmetry \cite{Uma} is generated by the out of equilibrium decay of the right handed, heavy Majorana neutrinos, which form an integral part of seesaw mechanism for neutrino masses and mixings. Since our proposal is model independent, we consider type I seesaw mechanism, just for simplicity.

\section{\textbf{Leptogenesis and Baryogenesis in Type I Seesaw SO(10) Models}}

In Grand Unified theories like SO(10), one right handed heavy Majorana neutrino per generation is added to Standard Model \cite{l,m,n,ibarra}, and they couple to left handed $ \nu $ via Dirac mass matrix $m_{D}$. When the neutrino mass matrix is diagonalised, we get two eigen values $ - $ light neutrino $ \sim $ $ \frac{m^{2}_{D}}{M_{R}} $ and a heavy neutrino state $ \sim $ $M_{R}$. This is called type I See Saw mechanism. Here, decay of the lightest of the three heavy RH Majorana neutrinos, $M_{1}$, i.e $M_{3}, M_{2}\gg M_{1}$ will contribute to $l-\bar{l}$ asymmetry (for leptogenesis), i.e $\epsilon^{CP}_{l}$. In the basis where RH $\nu$ mass matrix is diagonal, the type I contribution to $\epsilon^{CP}_{l}$ is given by decay of $M_{1}$
 \begin{equation}
\epsilon^{CP}_{l}=\frac{\Gamma(M_{1}\rightarrow lH) - \Gamma(M_{1}\rightarrow \bar{l} \bar{H})}{\Gamma(M_{1}\rightarrow lH) + (\Gamma(M_{1}\rightarrow \bar{l} \bar{H})}, 
\end{equation}
where $\Gamma(M_{1}\rightarrow lH)$ means decay rate of heavy Majorana RH $\nu$ of mass $M_{1}$ to a lepton and Higgs. We assume a 
normal mass hierarchy for heavy Majorana neutrinos. In this scenario the lightest of heavy Majorana neutrinos is in thermal equilibrium
 while the heavier neutrinos, $M_{2}$ and $ M_{3} $, decay. Any asymmetry produced by the out of equilibrium decay of $M_{2}$ and 
$ M_{3} $ will be washed away by the lepton number violating interactions mediated by $ M_{1} $. Therefore, the final lepton-antilepton asymmetry is given only by the CP-violating decay of $ M_{1} $ to standard model leptons (l) and Higgs (H). This contribution 
is \cite{Mh}:
\begin{equation}
\epsilon_{l}=-\frac{3M_{1}}{8\pi}\frac{Im[\Delta m^{2}_{\odot}R^{2}_{12}+\Delta m^{2}_{A}R^{2}_{13}]}{\upsilon^{2}\sum
|R_{1j}|^{2}m_{j}}. 
\end{equation}
where $\upsilon$ is the vev of the SM Higgs doublet that breaks the SM gauge group to $ U(1)_{em} $. $R$ is a complex orthogonal matrix with the property that $RR^{T} = 1$. $R$ can be parameterized as \cite{Osc}:
\begin{equation}
R = D_{\sqrt{M^{-1}}}Y_{\nu}UD_{\sqrt{K^{-1}}},
\end{equation}
where $Y_{\nu}$ is the matrix of neutrino Yukawa couplings. $ \upsilon $ is the vacuum expectation value of $\nu  $ Dirac masses. In the flavor basis, where the charged-lepton Yukawa matrix, $Y_{e}$ and gauge interactions are flavour-diagonal,
$ D_{K} = U^{T}KU  $, where $K=Y_{\nu}^{T}M_{R}^{-1}Y_{\nu}$. $U$ is the PMNS matrix and  $M_{R}$ is the RH neutrino Majorana scale. In the basis of right handed neutrinos,  $D_{M} = Diag(M_{1}, M_{2},M_{3})$ where $M_{3}, M_{2}\gg M_{1}$. Equation (8) relates the lepton asymmetry to both the solar ($\Delta m^{2}_{21}$) and atmospheric ($ \Delta m^{2}_{A} $) mass squared differences. 
Thus the magnitude of the matter-antimatter asymmetry can be predicted in terms of low energy oscillation parameters, $\Delta m^{2}_{21}$, $ \Delta m^{2}_{A} $ and a CPV phase. Here matrix $R$ is dependent on both $U_{PMNS}$ and $V_{CKM}$, and it can be shown that,
\begin{eqnarray*}
\text{Im}R^{2}_{13}& = &-\text{s}(2\delta_{q})\text{c}^{2}23_{l}\text{c}^{2}13_{l}\text{s}^{2}13_{q}-2\text{s}(\delta_{q})\text{c}{13}_{q}\text{c}{23}_{l}\text{c}^{2}13_{l}\text{s}12_{q}\text{s}13_{q}\text{s}23_{l}\\& &+2\text{s}(-\delta_{l}-\delta_{q})\text{c}12_{q}\text{c}13_{q}\text{c}23_{l}\text{c}13_{l}\text{s}13_{q}\text{s}13_{l}-2\text{s}(\delta_{l})\text{c}12_{q}\text{c}^{2}13_{q}\text{c}13_{l}\\& &\text{s}12_{q}\text{s}23_{l}\text{s}13_{l}-\text{s}(2\delta_{l})\text{c}^{2}12_{q}\text{c}^{2}13_{q}\text{s}^{2}13_{l}-2\text{s}(\delta_{l})\text{c}^{2}12_{q}\text{c}^{2}13_{q}\text{s}^{2}13_{l}
\end{eqnarray*}
\begin{eqnarray*}
\text{Im}R^{2}_{12}&=&2\text{s}(\delta_{q})\text{c}{13}_{q}\text{c}^{2}{12}_{l}\text{c}{23}_{l}\text{s}{12}_{q}\text{s}{13}_{q}\text{s}{23}_{l} +2\text{s}(\delta_{q})\text{c}{12}_{q}\text{c}{13}_{q}\text{c}{12}_{l}\text{c}{13}_{l}\text{s}{13}_{q}\text{s}{12}_{q} \\& &\text{s}{23}_{l}-\text{s}(2\delta_{q})\text{c}^{2}{12}_{l}\text{s}^{2}{13}_{q}\text{s}^{2}{23}_{l}-2\text{s}(\delta_{l}-\delta_{q})\text{c}{13}_{q}\text{c}{12}_{l}\text{c}^{2}{23}_{l}\text{s}{12}_{q}\text{s}{13}_{q}\\& &\text{s}{12}_{l}\text{s}{13}_{l}-2\text{s}(\delta_{l}-\delta_{q})\text{c}{12}_{q}\text{c}{13}_{q}\text{c}{23}_{l}\text{c}{13}_{l}\text{s}{13}_{q}\text{s}^{2}{12}_{l}\text{s}{23}_{l}-2\text{s}(\delta_{l})\text{c}^{2}{13}_{q}\\& &\text{c}{12}_{l}\text{c}{23_{l}}\text{s}^{2}{12}_{q}\text{s}{12}_{l}\text{s}{23}_{l}\text{s}{13}_{l}+2\text{s}(\delta_{l}-2\delta_{q})\text{c}{12}_{l}\text{c}{23}_{l}\text{s}^{2}{13}_{q}\text{s}{12}_{l}\text{s}{23}_{l}\text{s}{13}_{l}\\& &-2\text{s}(\delta_{l})\text{c}{12}_{q}\text{c}^{2}{13}_{q}\text{c}{13}_{l}\text{s}{12}_{q}\text{s}^{2}{12}^{l}\text{s}{23}_{l}\text{s}{13}_{l}+ 2\text{s}(\delta_{l}-\delta_{q})\text{c}{13}_{q}\text{c}{12}_{l}\text{s}{12}_{q}\\& &\text{s}{12}_{q}\text{s}{12}_{l}\text{s}^{2}{23}_{l}\text{s}{13}_{l}+ 2\text{s}(2\delta_{l}-2\delta_{q})\text{c}^{2}{23}_{l}\text{s}^{2}{13}_{q}\text{s}^{2}{12}_{l}\text{s}^{2}{13}_{l} +  2\text{s}(2\delta_{l}-\delta_{q}) \\& & \text{c}{13}_{q}\text{c}{23}_{l}\text{s}{12}_{q}\text{s}{13}_{q}\text{s}^{2}{12}_{l}\text{s}23_{l}\text{Sin}^{2}{13}_{l}+ \text{s}(2\delta_{l})\text{c}^{2}{13}_{q}\text{s}^{2}{12}_{q}\text{s}^{2}{12}_{l}\text{s}^{2}{23}_{l}\text{s}^{2}{13}_{l}
\end{eqnarray*}
\begin{eqnarray*}
R_{11}&=&\text{c}{12}_{q}\text{c}{13}_{q}\text{c}{12}_{l}\text{c}{13}_{l}+e^{-i\delta_{q}}\text{s}{13}_{q}\text{s}{12}_{l}\text{s}{23}_{l}-e^{-i\delta_{l}}e^{-i\delta_{q}}\text{s}{13}^{q}\text{c}{12}_{l}\text{c}{23}_{l}\text{s}{13}_{l} \\& &-\text{c}{13}_{q}\text{s}{12}_{q}\text{c}{23}_{l}\text{s}{12}_{l} - e^{-i\delta_{l}}\text{c}{13}_{q}\text{s}{12}_{q}\text{c}{12}_{l}\text{s}{23}_{l}\text{s}{13}_{l}
\end{eqnarray*}
\begin{eqnarray*}
R_{12}&=&\text{c}{12}_{q}\text{c}{13}_{q}\text{c}{13}_{l}\text{s}{12}_{l}-e^{-i\delta_{q}}\text{s}{13}_{q}\text{c}{12}_{l}\text{s}{23}_{l}-e^{-i\delta_{l}}e^{-i\delta_{q}}\text{c}{23}_{l}\text{s}{12}_{l}\text{s}{13}_{l}\\ & &\text{s}{13}_{q} -\text{c}{13}_{q}\text{s}{12}_{q}\text{c}{12}_{l}\text{c}{23}_{l} - e^{-i\delta_{l}}\text{c}{13}_{q}\text{s}{12}_{q}\text{s}{12}^{l}\text{s}{23}_{l}\text{s}{13}_{l}
\end{eqnarray*}
\begin{equation}
R_{13} = e^{-i\delta_{q}}\text{c}{23}_{l}\text{c}{13}_{l}\text{s}{13}_{q}-\text{c}{13}_{q}\text{c}{13}_{l}\text{s}{12}_{q}\text{s}{23}_{l} - e^{-i\delta_{l}}\text{c}{12}_{q}\text{c}{13}_{q}\text{s}{13}_{l}
\end{equation}
Here, $\text{ c}23_{l} $, $ \text{s}12_{l} $, $\text{c} 13_{l} $, etc represents the cosine of atmospheric mixing angle, sine of solar mixing angle and cosine of reactor mixing angle repectively. Similarly $ 23_{q} $, $ 12_{q} $, $ 13_{q} $ are the quark mixing angles. $\delta_{l}$ and $\delta_{q}$ are the leptonic CPV phase and quark CPV phase respectively. When left-right symmetry is broken at high intermediate mass scale $ M_{R} $ in SO(10) theory,
 CP asymmetry is given by
\begin{equation}
\epsilon_{l}=-\frac{3M_{1}}{8\pi}\frac{Im[\Delta m^{2}_{A}R^{2}_{13}]}{\upsilon^{2}\sum 
|R_{ij}|^{2}m_{j}} 
\end{equation}
where
$$|R_{11}|^{2}=\text{cos}^{2}(\theta_{12}^{l})\text{cos}^{2}(\theta_{13}^{l}), |R_{12}|^{2} = \text{sin}^{2}(\theta_{12}^{l})\text{cos}^{2}(\theta_{13}^{l}), |R_{13}|^{2} = \text{cos}^{2}(\delta_{l})\text{sin}^{2}(\theta_{13}^{l})+ \text{sin}^{2}(\delta_{l})\text{sin}^{2}(\theta_{13}^{l})$$and
\begin{equation}
\text{Im}R^{2}_{13} = -\text{sin}^{2}(2\delta_{l})\text{sin}^{2}(\theta_{13}^{l})
\end{equation}
The neutrino oscillation data used in our numerical calculations are summarised as follows {\cite{kol}}. 
$$\Delta m^{2}_{21}[10^{-5}eV^{2}] = 7.60 ^{+0.19}_{-0.18}$$ 
$$|\Delta m^{2}_{31}|[10^{-3}eV^{2}] = 2.48^{+0.05}_{-0.07}(2.38^{+0.05}_{-0.06})$$ 
$$\text{sin}^{2}\theta_{12} = 0.323{\pm 0.016}$$
$$\text{sin}^{2}\theta_{23} = 0.567^{+0.032}_{-0.124}(0.573^{+0.025}_{-0.039})$$ 
\begin{equation}
\text{sin}^{2}\theta_{13} = 0.0226{\pm 0.0012}(0.0229{\pm0.0012})
\end{equation}
For $\Delta m^{2}_{31}, sin^{2}\theta_{23}, sin^{2}\theta_{13}$, the quantities inside the bracket corresponds to inverted neutrino mass hierarchy and those outside the bracket corresponds to normal mass hierarchy. The errors are within the 1$\sigma$ range of the $\nu$ oscillation parameters. It may be noted that some results on neutrino masses and mixings using updated values of running quark and lepton masses in
SUSY SO(10) have also been presented in \cite{Gayatri}. Though we consider 3-flavour neutrino scenario, 4-flavour neutrinos with 
sterile neutrinos as fourth flavour, are also possible \cite{KBo}. It is worth mentionng that $ \nu $ masses and mixings can lead 
to charged lepton flavor violation in grand unified theories like SO(10) \cite{GG}.
\par
The origin of the baryon asymmetry in the universe (baryogenesis) is a very interesting topic of current research. A well known mechanism is the baryogenesis via leptogenesis, where the out-of-equilibrium decays of heavy right-handed Majorana neutrinos produce a lepton asymmetry which is transformed into a baryon asymmetry by electroweak sphaleron processes \cite{Hooft, Manton, Kuzmin}. Lepton asymmetry is partially converted to baryon asymmetry through B+L violating sphaleron interactions \cite{ME}. As proposed in \cite{SO}, a baryon 
asymmetry can be generated from a lepton asymmetry. The baryon asymmetry is defined as:
\begin{equation}
Y_{B} = \frac{n_{B}-n_{\bar{B}}}{s}= \frac{n_{B}-n_{\bar{B}}}{7n_{\gamma}}=\frac{\eta_{B}}{7}, 
\end{equation}
where $n_{B}, n_{\bar{B}},n_{\gamma}$ are number densities of baryons, antibaryons and photons respectively, $s$ is the entropy density,  $ \eta $ is the baryon-to-photon ratio, $ 5.7 \times 10^{-10} \leq \eta_{B} \leq 6.7 \times 10^{-10}$ (95 \%  C.L) \cite{B.D}. The lepton number is converted into the baryon number through electroweak sphaleron process \cite{Hooft,Manton,Kuzmin}.

\begin{equation}
Y_{B} = \frac{a}{a-1}Y_{L}, a = \frac{8N_{F} + 4N_{H}}{22N_{F}+13N_{H}},
\end{equation}
where $ N_{f} $ is the number of families and $ N_{H} $ is the
number of light Higgs doublets. In case of SM, $N_{f} = 3$ and $ N_{H} = 1 $. The lepton asymmetry is as follows:
\begin{equation}
Y_{L} = d\frac{\epsilon_{l}}{g^{*}}.
\end{equation}
$d$ is a dilution factor and $g^{*} = 106.75$ in the standard case \cite{SO}, is the effective number of light degrees of freedom in the theory. The dilution factor d \cite{SO} is, $d = \frac{0.24}{k(lnk)^{0.6}}$ for $k\geq 10 $ and $d = \frac{1}{2k}, d = 1$ for $1\leq k \leq 10$ and $0\leq k \leq 1$ respectively, where the parameter k \cite{SO} is, $ k=\frac{M_{P}}{1.7\upsilon^{2}32\pi\sqrt{g^{*}}}\frac{(M_{D}\dagger M_{D})_{11}}{M_{1}} $, here $ M_{P} $ is the Planck mass. We have used the form of Dirac neutrino mass matrix $M_{D}$ from \cite{Joshipura}.

\section{\textbf{Calculations, results and discussion}}
For the purpose of calculations, we use the current experimental data for three neutrino mixing angles as inputs, which are given at $1\sigma$ $-$ $3\sigma$ C.L, as presented in \cite{kol}. Here, we perform numerical analysis for both the hierarchies and octants. We explore the baryon asymmetry of the universe using Eq. (7)-Eq. (16) of the two hierarchies (NH and IH), two octants$-$ LO and HO, w ND, w/o ND (with and without near detector) and $ \delta_{CP} $ range at $ \geq $ 2$ \sigma $  over the corresponding distribution of $ \chi^{2} $-minima (for maximum sensitivity from Fig. 2(a), 2(b), for which the CP discovery potential of the DUNE is maximum). For our purpose, we shall carry out a general scanning of the parameters: $ \delta_{CP} $ range at $ \geq $ 2$ \sigma $ (from Fig. 2(a), 2(b)), $ \theta_{13} $ at its $ 3\sigma $ C.L and $ \Delta m^{2}_{31} $ at its $ 3\sigma $ C.L using the data given by the oscillation experiments \cite{Fgli, Frero, kol}. We scan the parameter space for IH, HO/LO in in the light of recent ratio of the baryon to photon density bounds, $ 5.7 \times 10^{-10} \leq \eta_{B} \leq 6.7 \times 10^{-10}$ (CMB) \cite{B.D} in the following ranges: 
$$ \delta_{CP}\hspace{.1cm} \in \hspace{.1cm} [0,\hspace{.1cm} 2\pi] $$
$$ \theta_{13}\hspace{.1cm} \in \hspace{.1cm} [7.8^{0},\hspace{.1cm}9.9^{0}] $$
\begin{equation}
\Delta m^{2}_{31} \hspace{.1cm} \in \hspace{.1cm} [-2.54*10^{-3},\hspace{.1cm}-2.20*10^{-3}]eV^{2} 
\end{equation}
Similarly constrained by the present BAU bounds we perform random scans for the following range of parameters in NH, HO/LO case:
$$ \delta_{CP}\hspace{.1cm} \in \hspace{.1cm} [0,\hspace{.1cm} 2\pi] $$
$$ \theta_{13}\hspace{.1cm} \in \hspace{.1cm} [7.7^{0},\hspace{.1cm}9.9^{0}] $$
\begin{equation}
\Delta m^{2}_{31} \hspace{.1cm} \in \hspace{.1cm} [2.30*10^{-3},\hspace{.1cm}2.65*10^{-3}]eV^{2} 
\end{equation}
We find that the updated BAU limit \cite{B.D} together with a large $\theta_{13}$ \cite{Fgli,Frero,Garc} puts significant constraints on the $ \delta_{CP} $-$ \theta_{13} $
parameter space in the IH, LO case. As can be seen from Fig. 3a, a part of the paramater space survives for $ \delta_{CP} \simeq 1.3-1.4 \pi $ in the IH, LO case as allowed by the current BAU constraint $ 5.7 \times 10^{-10} \leq \eta_{B} \leq 6.7 \times 10^{-10}$ (CMB), corresponding to $ \theta_{13} $ around  $ 8.9^{0}-9.5^{0}$. This leads to the conclusion that the parameter space for the best fit values of $ \delta_{CP} \simeq 1.3-1.4 \pi $ is allowed by the present BAU constraint. The allowed regions in Fig. 3a for the lower quadrant (0 to $\pi$)
requires $ \delta_{CP}$ spectra, i.e. $ \delta_{CP}$ to be equal to $ 88^{0} $ for $ \theta_{13} $ around  $ 9.09^{0} $ to $ 9.2^{0} $, $ 9.25^{0} $, $ 9.35^{0} $ to $ 9.5^{0} $, $ 9.6^{0} $ to $ 9.65^{0} $. Almost continuous values of $ \delta_{CP} $ ranging from $ 50^{0} $ to $ 140^{0} $ are allowed for $ \theta_{13} $, $ 9.3^{0} $ to $ 9.5^{0} $. For, $ \theta_{13} $ around $ 9.7^{0} $, the values of $ \delta_{CP} $ mostly favoured are $50^{0}, 69^{0}, 88^{0}, 90^{0}, 95^{0}, 99^{0}, 120^{0}, 130^{0}.$. The allowed region in the upper quadrant ($ \pi $ to 2$ \pi $) necessitates $ \delta_{CP} $ to be around $ 276.5^{0} $, for $\theta_{13}$ around $ 9.06^{0} $ to $ 9.12^{0} $, $ 9.3^{0} $ to $ 9.45^{0} $, $ 9.517^{0} $ as allowed by the current BAU bounds. Also $ \delta_{CP} =  290^{0} $ exists for $\theta_{13}$ around , $ 9^{0} $, $ 9.15^{0} $ to $ 9.3^{0} $, $ 9.35^{0} $ to $ 9.5^{0} $. Almost continuous $ \delta_{CP} $ ranging from $ 230^{0} $ to $ 340^{0} $ are allowed for $ \theta_{13} $ around $ 9.4^{0} $.

\begin{figure}[b]
\centerline{\begin{subfigure}[]{\includegraphics[width=8.0cm]{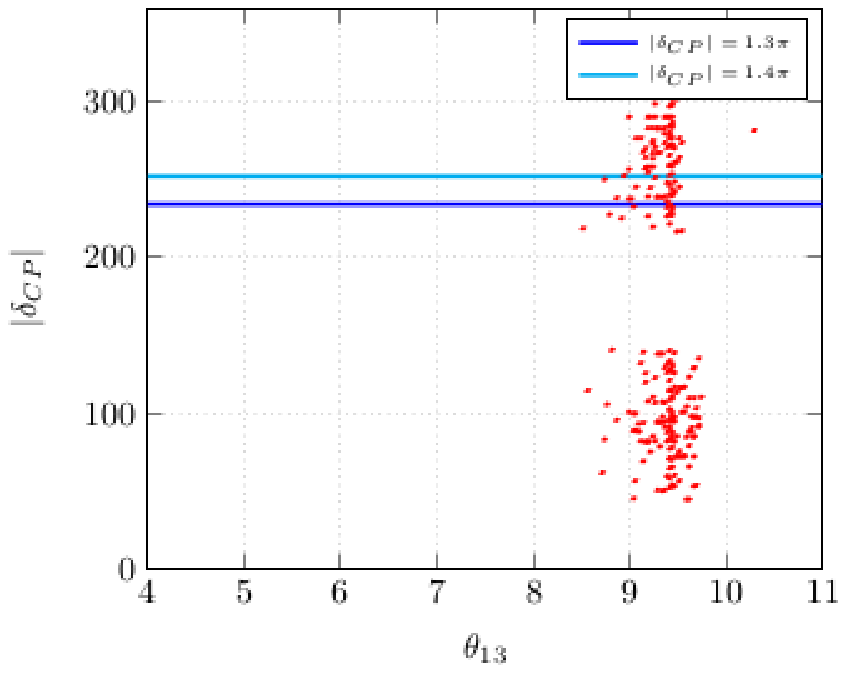}}\end{subfigure}
\begin{subfigure}[]{\includegraphics[width=8.0cm]{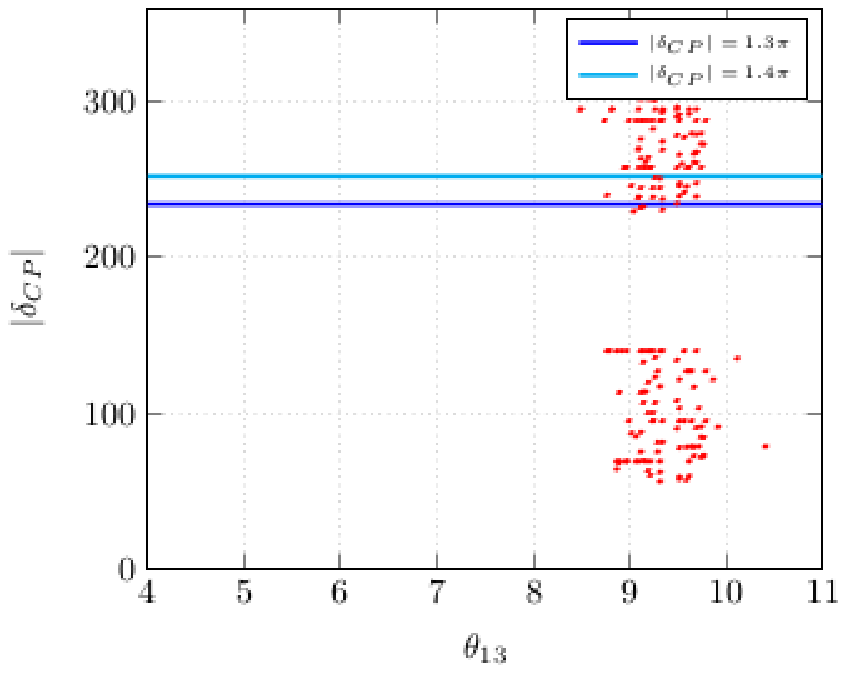}}\end{subfigure}}
\caption{Allowed region constrained by the present BAU bounds,  $5.7\times 10^{-10} < \eta_{B} < 6.7 \times 10^{-10} $ for $\delta_{CP}$, $\theta_{13}$ for the case when R matrix consists of both $V_{CKM}$ and $U_{PMNS}$. The regions are obtained by varying $\delta_{CP}$ range at $ \geq $ 2$ \sigma $ over the corresponding $ \chi^{2} $ minima distribution from fig. 2 and $ \theta_{13} $ with its experimental values varied within 3$ \sigma $. In Fig. 3a (3b) we show the plot for the IH, LO case (IH, HO case). The blue (cyan) horizontal line represents $ \delta_{CP} =1.3\pi (1.4\pi) $ around which the best fit values of CPV phase $ \delta_{CP}$ are assumed to lie.}
\end{figure}

For IH, HO case in Fig. 3(b), a part of  the $ \delta_{CP}-\theta_{13} $
parameter space exists for $ \delta_{CP} \simeq 1.3-1.4 \pi $ constrained by the current BAU limits for $ \theta_{13} $ around $ 9^{0}$, $ 9.097^{0}$, $ 9.1^{0}$, $ 9.2^{0}$, $ 9.3^{0}$, $ 9.5^{0}$, $ 9.55^{0}$, $9.6^{0} $ and $ 9.7^{0}$. Thus the parameter space for the best fit values of $ \delta_{CP} \simeq 1.3-1.4 \pi $ is allowed by the present BAU constraint \cite{B.D}. Fig. 3b reveals that the allowed regions for the lower quadrant (0 to $\pi$) requires $ \delta_{CP}$ to be around $ 69^{0} $,  for $ \theta_{13}$ $\sim$ $8.86^{0}-8.896^{0}$, $9.06^{0}-9.22^{0}$, $9.32^{0}$, $9.61^{0}$. There also exists $ \delta_{CP} = 88^{0}$ which constrains $ \theta_{13} $ to be around  $ 9.0^{0} $ to $ 9.1^{0} $. Also $ \delta_{CP} = 95^{0}$ suvives for $ \theta_{13} $ around  $ 8.99^{0} $, $ 9.24^{0} $, $ 9.352^{0} $, $ 9.5^{0}-9.64^{0} $ and $ 9.79^{0} $.  Almost continuous values of $ \delta_{CP} $ ranging from $ 50^{0} $ to $ 140^{0} $ are allowed for $ \theta_{13} $ varying from $ 8.75^{0} $ to $ 9.75^{0} $. In the upper quadrant ($ \pi $ to 2$ \pi $) for the present updated BAU constraint, the allowed region of $ \delta_{CP} $ parameter space becomes constrained with $ \delta_{CP} = 257.5^{0} $, for $ \theta_{13} $ around  $ 8.96^{0} $,  $ 9.0974^{0} $ to  $ 9.22^{0} $,  $ 9.5^{0} $,  $ 9.6^{0} $ and  $ 9.74^{0} $. Also the BAU constraints necessitates $ \delta_{CP} $ to be around $ 288^{0} $, for $\theta_{13}$ around  $ 9^{0} $, $ 9.06^{0} $ to $ 9.35^{0} $, $ 9.5^{0} $,  $ 9.69^{0} $,  $ 9.8 $ as allowed by the current BAU bounds. Also $ \delta_{CP} = 295^{0} $ survives for $\theta_{13}$ around  $ 9.09^{0} $, $ 9.26^{0} $, $ 9.34^{0} $, $ 9.5^{0} $,  $ 9.6^{0} $,  $ 9.69 $. Almost continuous $ \delta_{CP} $ ranging from $ 232^{0} $ to $ 340^{0} $ are allowed for $ \theta_{13} $ around $ 8.9^{0} $, $ 9.65^{0} $.
\par 
The constraints imposed on the $ \delta_{CP} $, $ \theta_{13} $ parameter in NH, HO/LO space are found to be more severe as compared to IH, HO/LO space. For, NH, LO only a particular value of CP violating phase, $ \delta_{CP} = 258.5^{0} $ corresponding to $ \theta_{13} = 9.02375$ is consistent with the BAU constraint. From our analysis we find that for NH, HO case we are unable to resolve the entanglement of the quadrant of $ \delta_{CP} $ and octant of $ \theta_{23} $  since no point in the parameter space ($ \delta_{CP}, \theta_{13} $) is in consistent with the recent ratio of baryon to photon density bounds,    
 $5.7\times 10^{-10} < \eta_{B} < 6.7 \times 10^{-10} $ . Therefore, this indicates that IH is the most favoured hierarchy for breaking the 4-fold degeneracy of Eq. (5), (6). All the  analysis presented above is for the case when R matrix consists of both $V_{CKM}$ and $U_{PMNS}$. 
\par 
No points in the $ \delta_{CP}-\theta_{13} $ parameter space, consistent with the BAU constraint, is able to break the entanglement of the quadrant of $ \delta_{CP} $ and octant of $ \theta_{23} $, when R matrix consists of $U_{PMNS}$ only.
\par 
In Fig. 4, 5 we display the allowed 3D-space ($ \delta_{CP} $,  $\theta_{13}$, $ \Delta m^{2}_{31} $) for breaking the 4-fold degeneracy of Eqn (5), (6) by varying the leptonic $ \delta_{CP} $ phase in all its possible range [0-2$ \pi $] at $ \geq 2 \sigma $ (from Fig. 2), the mixing parameter $ \theta_{13} $ within its $ 3\sigma $ level, and $ \Delta m^{2}_{31} $ at its 3$ \sigma $ C.L \cite{Frero}.

\begin{figure}[b]
\centerline{\includegraphics[width=16cm]{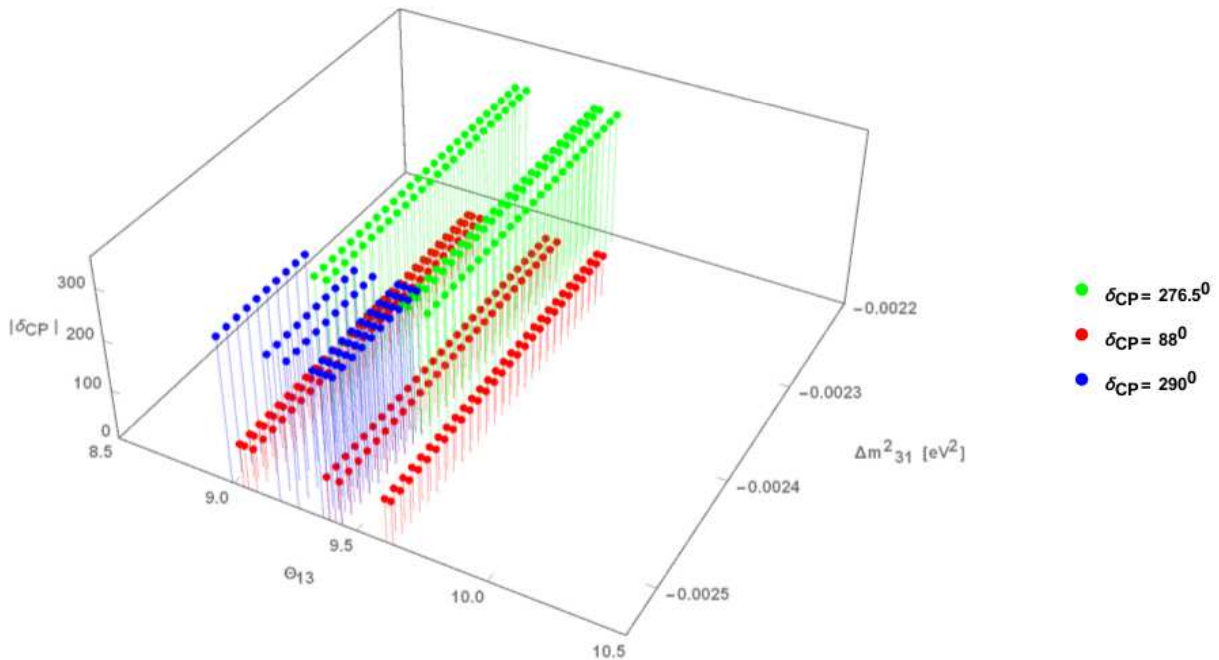}}
\caption{Allowed region constrained by the present BAU bounds,  $5.7\times 10^{-10} < \eta_{B} < 6.7 \times 10^{-10} $ for $\delta_{CP}$, $\theta_{13}$ and $ \Delta m^{2}_{31} $ for the case when R matrix consists of both $V_{CKM}$ and $U_{PMNS}$. The regions are obtained by varying $\delta_{CP}$ range at $ \geq $ 2$ \sigma $ over the corresponding $ \chi^{2} $ minima distribution from fig. 2 and $ \theta_{13} $ with its experimental values varied within 3$ \sigma $ and $ \Delta m^{2}_{31} $ at its 3$ \sigma $ C.L. The results of our calculation are presented for  IH, LO case.}
\end{figure}

From Figure 4 one can easily see the favoured values of $ \delta_{CP} $, $ \theta_{13} $ and $ \Delta m^{2}_{31} $ for IH, LO case, allowed by the updated recent ratio of photon density to baryon density bounds, $5.7\times 10^{-10} < \eta_{B} < 6.7 \times 10^{-10} $ ( shown in Table I).

\begin{table}[H]
\begin{center}
\begin{tabular}{|c|c|c|c|c|}

\hline 
\textbf{S.No}& \textbf{Leptonic CPV Phase $\delta_{CP}$}  & \textbf{$[\Delta m^{2}_{31}]eV^{2}$}& \textbf{$ \theta_{13}$}& \textbf{Quadrant of $\delta_{CP}$ }\\ 
\hline 
$1$&$\delta_{CP} = 88^{0}$&$ [-2.54*10^{-3},\hspace{.1cm}-2.21*10^{-3}]$ & $ 9.0417^{0},9.0697^{0} ,  9.0974^{0},$  & I \\ & & & $ 9.3917^{0},  9.4417^{0} ,  9.6167^{0},$ & \\& & &  $ 9.6417^{0} $  & \\
\hline
$2$ &  $ \delta_{CP} = 276.5^{0}$ & $ [-2.45*10^{-3},\hspace{.1cm}-2.21*10^{-3}]$ &$  9.0667^{0} ,  9.0974^{0} ,$ & IV \\ & & &  $9.3167^{0}- 9.4417^{0} ,  9.5167^{0} $ & \\
\hline
$3$& $ \delta_{CP} = 290^{0}$ &$  [-2.42*10^{-3},\hspace{.1cm}-2.54*10^{-3}]$& $ 9^{0} ,  9.1817^{0} , 9.2667^{0},$ & IV \\ & & & $ 9.3667^{0}- 9.4417^{0}$ & \\
\hline

\end{tabular}
\end{center}
\caption{The summary of our calculated values of $ \delta_{CP} $, $ \theta_{13} $ and $ \Delta m^{2}_{31} $   in case of IH, LO for $ R_{1j} $  elements of R Matrix comprising of  $ U_{PMNS} $ and $ V_{CKM}$.}
\end{table}

For IH, HO case, the results of
our numerical analysis are shown in Fig. 5 which shows allowed  ($ \delta_{CP} $, $ \theta_{13} $, $ \Delta m^{2}_{31} $) space as allowed by the current BAU bounds. The values of $ \delta_{CP} $, $ \theta_{13} $ and $ \Delta m^{2}_{31} $ which are favoured simultaneously in consistent with $\eta_{B}$ constraints, $5.7\times 10^{-10} < \eta_{B} < 6.7 \times 10^{-10} $ \cite{B.D}, are as shown in Table II. For NH case we get only one point as shown in Eq. (19).

\begin{equation}
\text{NH, HO,} \hspace{.1cm} \text{III} \hspace{.1cm} \text{ quadrant } \hspace{.1cm}\text{ of} \hspace{.1cm}\text{ Leptonic }\hspace{.1cm} \delta_{CP}  \hspace{.1cm} \text{ phase}\hspace{.1cm} \delta_{CP}=258.4^{0} or \hspace{.1cm}  1.436\pi
\end{equation}

\begin{table}[H]
\begin{center}
\begin{tabular}{|c|c|c|c|c|}

\hline 
\textbf{S.No}& \textbf{$\delta_{CP}$}  & \textbf{$[\Delta m^{2}_{31}]eV^{2}$}& \textbf{$ \theta_{13}$} & \textbf{Quadrant of $\delta_{CP}$ }\\ 
\hline 
$1$& $ 95^{0}$ &$[-2.54*10^{-3},\hspace{.1cm}-2.21*10^{-3}]$ & $8.9917^{0} ,  9.2417^{0} ,  9.352^{0},$ & II \\ & & &  $9.5167^{0}, 9.567^{0}, 9.6417^{0},$ & \\ & & & $ 9.7917^{0}  $  & \\
\hline
$2$&$ 69^{0}$ &$ [-2.54*10^{-3},\hspace{.1cm}-2.38*10^{-3}]$ & $ 8.8667^{0}, 8.8917^{0} ,  8.9667^{0},$ & I \\ &  & & $9.0667^{0}, 9.0917^{0}, 9.1667^{0},$ & \\ &  & & $9.2167^{0}, 9.3167^{0} , 9.6167^{0} $  & \\
\hline
$3$ &$140^{0}$ & $[-2.54*10^{-3},\hspace{.1cm}-2.21*10^{-3}]$ & $ 8.7667^{0} , 8.8667^{0}, 8.9167^{0},$ & II \\ &  & & $ 8.9667^{0},9.1167^{0}, 9.1417^{0},$ & \\ &  & &$9.1917^{0}-9.2417^{0}, 9.3167^{0},$ & \\ &  & &$9.3417^{0}, 9.5667^{0}, 9.6917^{0}$  & \\
\hline
$4$& $ 257.5^{0}$ & $  [-2.54*10^{-3},\hspace{.1cm}-2.21*10^{-3}]$ &   $8.9667^{0}, 9.0974^{0}, 9.1417^{0},$ & III \\ &  & & $9.1667^{0}, 9.2167^{0}, 9.4917^{0},$
&\\ &  & &
$ 9.5917^{0}, 9.6167^{0}, 9.7417^{0}$ & \\
\hline
$5$&$ 295^{0}$ & $[-2.3*10^{-3},\hspace{.1cm}-2.21*10^{-3}]$&  $8.4917^{0}, 8.8167^{0}, 9.0917^{0},$ & IV\\ &  & & $9.2667^{0}, 9.4917^{0}, 9.6167^{0}$   & \\
\hline
$6$& $288^{0}$ & $  [-2.44*10^{-3},\hspace{.1cm}-2.28*10^{-3}]$ & $ 8.7817^{0}, 8.9917^{0}, 9.0667^{0},$ & IV \\ &  & & $ 9.1167^{0},9.1417^{0}, 9.1917^{0},$ 
& \\ &  & &
$ 9.2417^{0}, 9.3167^{0}, 9.3417^{0},$& \\ &  & & $9.5167^{0}, 9.6917^{0}, 9.7917^{0}$ & \\
\hline
\end{tabular}
\end{center}
\caption{The summary of our calculated values of $ \delta_{CP} $, $ \theta_{13} $ and $ \Delta m^{2}_{31} $   in case of IH, HO for $ R_{1j} $  elements of R Matrix comprising of  $ U_{PMNS} $ and $ V_{CKM}$.}
\end{table}

\begin{figure}[b]
\centerline{\includegraphics[width=16cm]{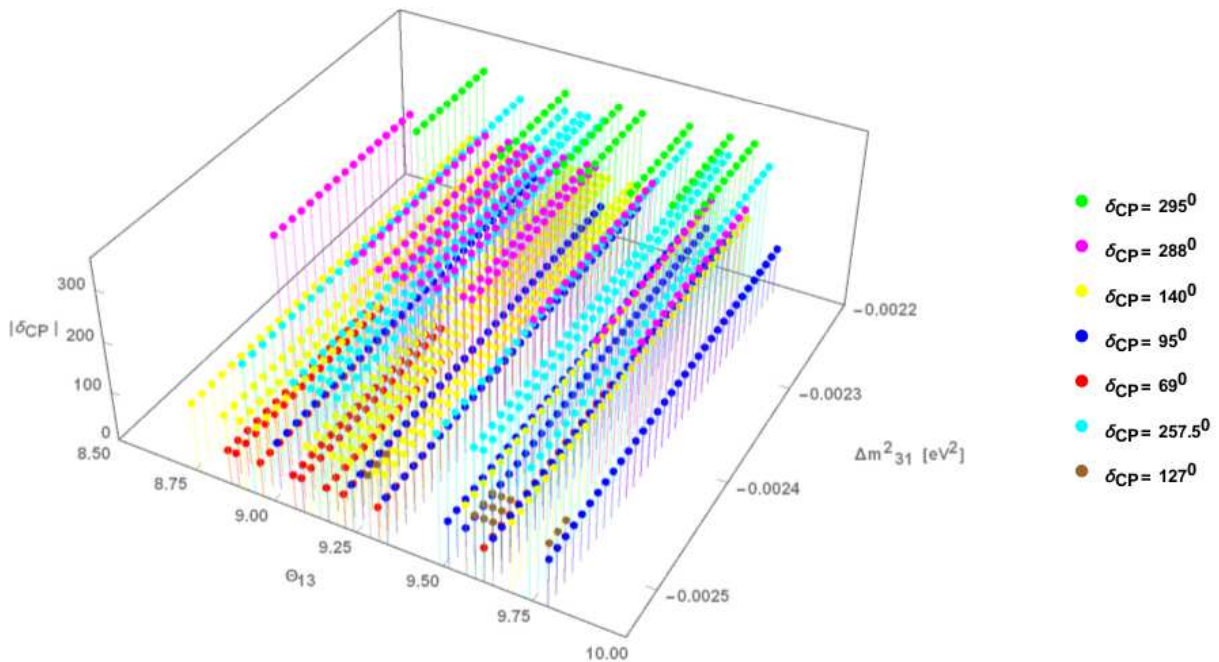}}
\caption{Allowed 3-D region constrained by the present BAU bounds,  $5.7\times 10^{-10} < \eta_{B} < 6.7 \times 10^{-10} $ for $\delta_{CP}$, $\theta_{13}$ and $ \Delta m^{2}_{31} $ for the case when R matrix consists of both $V_{CKM}$ and $U_{PMNS}$. The regions are obtained by varying $\delta_{CP}$ range at $ \geq $ 2$ \sigma $ over the corresponding $ \chi^{2} $ minima distribution from fig. 2, $ \theta_{13} $ with its experimental values varied within 3$ \sigma $ from its central values and $ \Delta m^{2}_{31} $ at its 3$ \sigma $ C.L. The results of our calculation are presented for  IH, HO case.}
\end{figure}

\section{\textbf{Conclusion}}
Measuring CP violation in the lepton sector is one of the most challenging tasks today. A systematic study of the CP sensitivity of the current and upcoming LBNE/DUNE is done in our earlier work \cite{DK} which may help a precision measurement of leptonic $ \delta_{CP}$ phase. In this work, we studied how the entanglement of the quadrant of leptonic CPV phase and octant of atmospheric mixing angle $ \theta_{23} $ at LBNE/DUNE, can be broken via leptogenesis and baryogenesis. Here, we have considered the effect of ND only in LBNE, on sensitivity of CPV phase measurement, but similar conclusions would hold for the effect of reactor experiments as well. This study is done for both
 the octants and hierarchies. We considered two cases of fermion rotation matrix - PMNS only, and CKM+PMNS.
Following the results of \cite{DK}, the enhancement of CPV sensitivity with respect to its quadrant is utilized here to calculate the values of lepton-antilepton symmetry. Then, this is used to calculate the  value of BAU. This is an era of precision measurements in neutrino physics. We therefore considered variation of $\Delta m^{2}_{31}$ and $\theta_{13}$ within its 3$ \sigma $ range from their central values. We calculated baryon to photon ratio, and compared with its experimentally known best fit value. 
\par 
We have made a complete numerical analysis of the 3 dimensional parameters, $\delta_{CP}$, $\theta_{13}$ and $\Delta m^{2}_{31}$ that encode the breaking of the entanglement of the quadrant of CPV phase and Octant of $\theta_{23}$ in presence of the latest constraints on $ |\eta_{B}| $, $5.7\times 10^{-10} < \eta_{B} < 6.7 \times 10^{-10} $, by taking neutrino oscillation mixings and mass scales as indicated by the experiments. By allowing $\delta_{CP}$ range to vary within $ [ 0-2\pi]  $ interval at $ \geq 2\sigma $ over the $ \chi^{2}- $ minima distribution from Fig. 2, we have studied the absolute values of both $ \theta_{13} $, $ \delta_{CP} $ parameters in order to  break the 4-fold degeneracy of Eq. (5), (6).
\par 
The current data shows a preference of $\delta_{CP}$ towards $ 1.5\pi $. From our analysis, one of the leptonic CPV phase determined in IH, LO of $ \theta_{23} $ case is $\delta_{CP}=276.5^{0}$ or $ 1.536\pi $ corresponding to $\theta_{13} \sim 9.0667^{0} ,  9.0974^{0} ,  9.3167^{0}- 9.4417^{0} ,  9.5167^{0} $ and $\Delta m^{2}_{31} \hspace{.1cm} \in \hspace{.1cm} [-2.45*10^{-3},\hspace{.1cm}-2.21*10^{-3}]eV^{2}$ which is near to the preferred data $\delta_{CP}=\frac{3}{2}\pi$ reported in \cite{Abe, Bian, S}. The current analysis reveals that our calculated value of $\delta_{CP}= 1.43\pi$ of IH, HO of $ \theta_{23} $ case is close to the best fit value of $\delta_{CP}= 1.48\pi$ for inverted ordering from global fit results \cite{kol} and also $\delta_{CP}= 1.436\pi$ of NH, HO of $ \theta_{23} $ case is favoured with the recent hint of $\delta_{CP}= 1.41\pi$ for normal hierarchy \cite{kol}.
\par 
The main results of this work are presented in Table I and II and Eq. (19) which show that leptonic CPV phase in all the four quadrants are allowed which lie within the constraints of present BAU. These values also contain the best fit values of leptonic CPV phase as discussed earlier.
\par 
These results could be important, as the quadrant of leptonic CPV phase, and octant of atmospheric mixing angle $\theta_{23}$ are yet not fixed experimentally. Also, they are significant in context of precision measurements of neutrino oscillation parameters, specially the leptonic CPV phase, $ \Delta m^{2}_{31} $ and the reactor angle $\theta_{13}$.

 \par 
Future experiments like DUNE/LBNEs and Hyper-Kamionande \cite{Mth} looking for the leptonic CPV phase $ \delta_{CP} $ together with an improvement in the precision determination on the mixing angles would certainly provide worthy informations to support or rule out the scenario presented in this work for breaking the entanglement of quadrant of CPV phase and Octant of $ \theta_{23} $  .

\section*{Acknowledgments}
GG would like to thank UGC, India, for providing RFSMS fellowship to her, during which  this work was done. KB thanks DST-SERB, Govt of India, for financial support through a project.


\section*{References}

\begin{thebibliography}{20}    
\bibitem{Fgli} G. Fogli, E. Lisi, A. Marrone, D. Montanino, A. Palazzo, et al., Phys. Rev. \textbf{D86}, 013012 (2012). arxiv:1205.5254 [hep-ph]; F. P. An et al. [Daya Bay Collaboration], Phys. Rev. Lett, \textbf{108}, 171803 (2012); Chin. Phys.C \textbf{37}, 011001 (2013), arXiv:1210.6327 [hep-ex]; J. K. Ahn et al. [RENO Collaboration], Phys. Rev. Lett. \textbf{108}, 191802 (2012). arXiv:1204.0626 [hep-ex]; Y. Abe et al. [Double Chooz Collaboration] Phys. Rev. Lett., \textbf{108}, 131801 (2012).
\bibitem{Frero} D. Forero, M. Tortola, and J. Valle, Phys Rev. \textbf{D86}, 073012 (2012). arxiv:1205.4018 [hep-ph]
\bibitem{Garc} M. Gonzalez-Garcia, M. Maltoni, J. Salvado, and T. Schwetz, JHEP 1212, 123 (2012). arxiv:1209.3023 [hep-ph]
\bibitem{DK} Debajyoti Dutta, Kalpana Bora, Mod. Phys. Lett. \textbf{A30}(07), 1550017, (2015). arxiv:1409.8248
\bibitem{MG} Monojit Ghosh, Pomita Ghoshal, Srubabati Goswami, Sushant K. Raut, Nucl. Phys. \textbf{B884}, 274-304 (2014). arxiv:1401.7243
\bibitem{PT} I. Girardi, S. T. Petcov, A.V. Titov, Eur. Phys. J. \textbf{C75}(7), 345 (2015). arxiv:1504.00658.
\bibitem{Kang} Sin Kyu Kang, M Tanimoto, Phys. Rev. \textbf{D91}(7), 073010 (2015). arxiv:1411.3104.
\bibitem{LHCb} LHCb Collaboration (Roel Aaij (NIKHEF, Amsterdam) et al.) Phys. Rev. Lett. \textbf{114}, 041801, 4 (2015), arxiv:1504.00658. LHCB-PAPER-2014-059, CERN-PH-EP-2014-271, LHCB-PAPER-2014-059-AND-CERN-PH-EP-2014-271
\bibitem{Patrik} Patrick Huber, Manfred Lindner, Thomas Schwetz, Walter Winter, JHEP \textbf{0911} 044 (2009). arXiv:0907.1896
\bibitem{KD} Kalpana Bora, Debajyoti Dutta, Pomita Ghoshal,  Mod. Phys. Lett. \textbf{A30}(14), 1550066, (2015). arxiv:1405.7482 
\bibitem{Gonzalez} M. C. Gonzalez-Garcia, M. Maltoni, A. Yu. Smirnov, Phys. Rev.\textbf{ D70}, 093005 (2004). hep-ph/0408170
\bibitem{Animesh} Animesh Chatterjee, Pomita Ghoshal, Srubabati Goswami, Sushant K.Raut, JHEP \textbf{1306}, 010(2013). arxiv:1302.1370 
\bibitem{Choubey} Sandhya Choubey, Anushree Ghosh, JHEP 1311, 166 (2013). arxiv:1309.5760
\bibitem{Daljeet} Daljeet Kaur, Naimuddin, Sanjeev Kumar, Eur. Phys. J.\textbf{ C75}(4), 156(2014). arxiv:1302.1370 
\bibitem{LBNE} LBNE Collaboration, C. Adams et al., arXiv:1307.7335.
\bibitem{Akiri} T. Akiri et al. [LBNE Collaboration], arXiv:1110.6249 [hep- ex]
\bibitem{Ayres} NO$\nu$A Collaboration, D. Ayres et al., hep-ex/0503053.
\bibitem{T2K} T2K Collaboration, K. Abe et al., Phys. Rev. Lett. \textbf{107}, 041801 (2011). arXiv:1106.2822
\bibitem{Minos} MINOS Collaboration, P. Adamson et al., Phys. Rev. Lett. \textbf{107} 181802 (2011). arXiv:1108.0015
\bibitem{DA} D. Autiero, J. Aysto, A. Badertscher, L. B. Bezrukov, J. Bouchez, et al., JCAP \textbf{0711}, 011 (2007). arxiv:0705.0116  
\bibitem{Branco}G. C. Branco, R. G. Felipe and F. R. Joaquim, Rev. Mod. Phys. \textbf{84}, 515 (2012). arXiv:1111.5332 [hep-ph].
\bibitem{Pcarvo} B. Pontecorvo. Sov. Phys. JETP, 6, 429, 1957 ; Sov. Phys. JETP, 26, 984, (1968) ; Z. Maki, M. Nakagawa and S. Sakata, Prog. Theor. Phys., 28, 870 (1962).
\bibitem{Capp} F. Capozzi, G.L. Fogli, E. Lisi, A. Marrone, D. Montanino, A. Palazzo , Phys. Rev. \textbf{D89}, 093018 (2014), arXiv:1312.2878 [hep-ph]; G. L. Fogli, E. Lisi, A. Marrone, D. Montanino, A. Palazzo and A. M. Rotunno, Phys. Rev.\textbf{ D 86}, 013012 (2012), arXiv:1205.5254 [hep-ph]; D. V. Forero, M. Tortola and J. W. F. Valle, Phys. Rev. D 90 (2014), arXiv:1405.7540 [hep-ph].
\bibitem{Gon} M. C. Gonzalez-Garcia, M. Maltoni and T. Schwetz, JHEP \textbf{1411}, 052 (2014), arXiv:1409.5439 [hep-ph].
\bibitem{Varger} V. Barger, D. Marfatia, and K. Whisnant, Phys. Rev. \textbf{D66}, 053007 hep-ph/0206038.
\bibitem{Minakata} H. Minakata and H. Nunokawa, JHEP \textbf{0110}, 001 (2001). hep-ph/0108085
\bibitem{kol} L . M. Cebola, D. E. Costa, R. G. Felipe, Eur. Phys. J. C \textbf{76} 3,156 (2016), arXiv:1601.06150; K. A. Olive et al. [Particle Data Group Collaboration], Chin. Phys. \textbf{C 38}, 090001 (2014); D. V. Forero, M. Tortola and J. W. F. Valle, Neutrino oscillations refitted, Phys. Rev. \textbf{D 90} (2014) 093006, [1405.7540]; G.L. Fogli, E. Lisi, A. Marrone, D. Montanino, A. Palazzo, and A.M. Rotunno, Phys. Rev.\textbf{ D
86}, 013012 (2012); M.C. Gonzalez-Garcia, M. Maltoni, J. Salvado, and T. Schwetz, JHEP 1212, 123 (2012); F. Capozzi, G. L. Fogli, E. Lisi, A. Marrone, D. Montanino and A. Palazzo, arXiv:1312.2878 [hep-ph].
\bibitem{Abe} K. Abe et al. (T2K), Measurements of neutrino oscillation in appearance and disappearance channels by the T2K experiment with 6.610 20 protons on target, Phys. Rev. D91 (2015) 7 072010, arXiv:1502.01550 [hep-ex].
\bibitem{Bian} J. Bian, First Results of ν e Appearance Analysis and Electron Neutrino Identification at NOvA, in Meeting of the APS
Division of Particles and Fields (DPF 2015) Ann Arbor, Michigan, USA, August 4-8, 2015 (2015) arXiv:1510.05708
[hep-ex], URL \textit{http://inspirehep.net/record/1399048/files/}arXiv:1510.05708.pdf.
\bibitem{S} URL
\textit{https://indico.cern.ch/event/361123/session/2/contribution/348/attachments} /1136004/1625868/SK-atmospheric-kachulis-dpf2015.pdf.
\bibitem{Bari}W. Buchmuller, P. Di Bari (DESY), M. Plumacher, Nucl.Phys. B\textbf{643}, 367-390 (2002), Nucl.Phys. B\textbf{793}, 362 (2008), hep-ph/0205349
\bibitem{RM} R.N. Mohapatra, Hai-Bo Yu, Phys.Lett. \textbf{B644}, 346-351 (2007), hep-ph/0610023
\bibitem{Bhupal} S. Bhupal Dev. talk presented at DAE HEP symposium, IITG, Dec 8-12, 2014.
\bibitem{Pd} P.S. Bhupal Dev, Chang-Hun Lee, R.N. Mohapatra, J. Phys. Conf. Ser. \textbf{631}1, 012007 (2015) 
\bibitem{GC} Gayatri Ghosh, Kalpana Bora, talk presented at DAE HEP symposium, IITG, Dec 8-12, 2014,  Springer Proc. Phys. \textbf{174} (2016) 287-291.
\bibitem{Mki}Z. Maki, M. Nakagawa and S. Sakata, Prog. Theor. Phys. \textbf{28}, 870 (1962).
\bibitem{Uma}Narendra Sahu, S.Uma Sankar, Phys.Rev. \textbf{D71},  013006  (2005), hep-ph/0406065 
\bibitem{l}M. Fukugita and T. Yanagida, Phys. Lett. \textbf{B17}4, 45 (1986).
\bibitem{m}R. N. Mohapatra and X. Zhang; Phys. Rev. \textbf{D46}, 5331 (1992).
\bibitem{n} M. Plumacher, Z. Phys. \textbf{74}, 549 (1997).
\bibitem{ibarra} S. Davidson and A. Ibarra, Phys. Lett.\textbf{ B 535}, 25 (2002). arXiv: hep-ph/0202239
\bibitem{Mh}R. N. Mohapatra, S. Nasri, Hai-Bo Yu, Phys.Lett. \textbf{B615}, 231-239 (2005). hep-ph/0502026
\bibitem{Osc}J.A. Casas and A. Ibarra, Nucl. Phys. B618, 171-204 (2001)hep-ph/0103065
\bibitem{Gayatri} Kalpana Bora, Gayatri Ghosh, J. Phys. Conf. Ser., \textbf{481}, 012016 (2014). 
\bibitem{KBo} Kalpana Bora, Debajyoti Dutta, Pomita Ghoshal, JHEP  \textbf{12}, 025 (2012). arXiv:1206.2172
\bibitem{GG} Kalpana Bora, Gayatri Ghosh, Eur. Phys. J. C\textbf{75}, (2015) 9, 428, arxiv:1410.1265
\bibitem{Hooft} G. t Hooft, Phys. Rev. Lett \textbf{37}, 8 (1976)
\bibitem{Manton} F. R. Klinkhamer, N. S. Manton, Phys. Rev. D \textbf{30}, 2212 (1984)
\bibitem{Kuzmin}V. A. Kuzmin, V. A. Rubakov, M.E shaposhikov, Phys Lett B \textbf{155}, 36 (1985)
\bibitem{ME} S. Yu Khlebnikov, M.E Shaposhnikov, Nucl phys B \textbf{3D8}, 885 (1968)
\bibitem{SO} F. Buccella, D. Falcone, F. Tramontano, Physics Letters B \textbf{524} 241–244 (2002).hep-ph/0108172
\bibitem{B.D} B. D. Fields, P. Molarto and S. Sarkar, "Big Bang Nucleosynthesis," in Review of PDG-2014 (Astrophysical Constants and Parameters)
\bibitem{Joshipura} Anjan S. Joshipura, Ketan M. Patel, Phys. Rev. \textbf{D83}, 095002(2011), arxiv:1102.5148
\bibitem{Mth} Matthew Malek, talk presented at 17th Lomonoscov Conference on Elementary Particle Physics, Moscow State University- 26 August 2015.
\end{thebibliography}
\end{document}